# Interfacial Tension Modulation of Liquid Metal via Electrochemical Oxidation

*Minyung Song[1], Karen E. Daniels[2], Abolfazl Kiani[1] Sahar Rashidnadimi[1] and Michael D. Dickey[1]\**

1 Department of Chemicals & Biomolecular Engineering, North Carolina State University, Raleigh, NC 27595
2 Department of Physics, North Carolina State University, Raleigh, NC 27595

**Abstract**

This progress report summarizes recent studies of electrochemical oxidation to modulate the interfacial tension of gallium-based alloys. These alloys, which are liquid at ambient conditions, have the largest interfacial tension of any liquid at room temperature. The ability to modulate the tension offers the possibility to create forces that change the shape and position of the metal. It has been known since the late 1800s that electrocapillarity–the use of potential to modulate the electric double layer on the surface of metals in electrolyte–lowers the interfacial tension of liquid metal. Yet, this phenomenon can only achieve modest changes in interfacial tension since it is limited to potential windows that avoid reactions. A recent discovery suggests that reactions driven by the electrochemical oxidation of gallium alloys cause the interfacial tension to decrease from ~500 mN/m at 0 V to ~0 mN/m at ~0.8 V, a change in tension that goes well-beyond what is possible via conventional electrocapillarity or surfactants. The changes in tension are reversible; reductive potentials return the metal back to a state of high interfacial tension. This report aims to summarize key work and introduce beginners to this field by including electrochemistry basics while addressing misconceptions. We discuss applications that utilize modulations in interfacial tension of liquid metal and conclude with remaining opportunities and challenges that need further investigation.

**What are liquid metals?**

Among all of the elements in the periodic table, there are only five metals that are liquid near room temperature: francium (Fr), cesium (Cs), rubidium (Rb), mercury (Hg), and gallium (Ga). However, the first three elements are not practical for applications because they are either radioactive (Fr, Cs) or explosively reactive with air (Rb). Mercury has been widely used in the past but is now avoided due to its toxicity [1], [2]. By process of elimination, gallium and gallium-based alloys are the safest liquid metals [3].

Gallium, discovered in 1875, has an extremely low vapor pressure even at high temperature (effectively zero at room temperature and only 1kPa at 1037°C [4], whereas water has a vapor pressure of 1kPa at 7°C). This low vapor pressure keeps the liquid from evaporating and eliminates potential danger of vapor inhalation. While gallium should be found in its solid phase at room temperature because its melting point is 30°C, it is often found as a liquid due to its ability to supercool. To assure a room-temperate liquid phase, eutectic alloys containing at least one additional metal are commonly used to lower the melting point below room temperature. Two popular commercial alloys are Galinstan (68.5 wt.% Ga, 21.5 wt.% In, 10.0 wt.% Sn) and EGaIn (75.5 wt.% Ga, 24.5 wt.% In), for which the melting point is 10.7°C and 15.5°C, respectively [5]. In air, these alloys all form surface oxides within microseconds due to a high reactivity of gallium with oxygen. Because Ga oxidizes more readily than the other components in these alloys, the surface oxides are dominated by gallium oxide species and these alloys thereby exhibit similar interfacial behavior to gallium in air [6]; thus, we discuss these gallium-based alloys without distinguishing them by their bulk composition except where warranted.

**Interfacial tension modulation**

Metals have significantly larger surface tension (>400 mN/m depending on the liquid metal [7]) than common fluids such as water (72 mN/m). The large surface tension of metals is due to the metallic bonds [4]. Gallium is notable for having the highest surface tension (708mN/m) of any liquid in air near room temperature [8], [9] and alloys of Ga, such as EGaIn, also have enormous surface tension (624 mN/m) [10].

We note that the term *surface tension* refers to the tension of fluid interfaces in contact only with a vapor phase, whereas *interfacial tension* refers more broadly to the tension of fluids in contact with other materials. Therefore, *interfacial tension* is more appropriate to use in this context. Note that the presence of oxides at the fluid-fluid interface creates additional effects, to be discussed below, which means this would most properly be called an *effective interfacial tension.*

Techniques for on-demand manipulation of the effective interfacial tension of liquid metals are highly desirable for applications involving liquid metal at the sub-mm length scale, where interfacial forces dominate. For example, Laplace pressure scales with $\gamma R^{-1}$, where $\gamma$ is the interfacial tension and $R$ is the characteristic dimension of the liquid metal (e.g. the radius of a droplet or wire). Thus, controlling $\gamma$ is a powerful way to generate forces on small volumes of fluid. Using such forces and combining with key properties of liquid metals such as their low viscosity [11], high thermal and electrical conductivity [12], it is possible manipulate liquid metals into different positions and shapes [13]–[16], and even create devices such as pumps [17], valves [18], and RF devices [19] by manipulating liquid metals inside microfluidic channels [11], [20]–[23].

There are two existing ways to lower the interfacial tension of liquids—surfactants (surface active agents, molecules that concentrate at interfaces) and electrical potential—with the former being the most common since it does not require a conductive fluid. For example, adding a common surfactant—sodium dodecyl sulfate (SDS)—to water lowers its surface tension. The surface tension decreases with the concentration of surfactant until reaching the critical micelle concentration (CMC), after which the surface saturates with surfactant and the surface tension plateaus. In the case of SDS, the surface tension of water decreases from 72 mN/m (for a clean surface) down to 32 mN/m [24] at the CMC. By combining other surfactants with SDS, the value can drop further [25]. Fluorosurfactants are among the most effective compounds to lower the surface tension of aqueous solution, to a value of 15mN/m [26]. However, surfactants only bring about modest changes in interfacial tension in liquid metals. For example, adding an alkythiol monolayer to Hg in the vapor phase lowers the surface tension by ~60 mN/m, similar in magnitude to what surfactants do for water [27], [28], but here accounting for only a 10% change due to the large surface tension of clean Hg. Polymers, such as polyvinyl alcohol, have been used on gallium-based alloys to help stabilize colloidal suspensions [12], [29]. However, we know of no reports addressing changes in effective interfacial tension of those gallium-based alloys using surfactants because the presence of oxide complicates the effects of surfactants.

While typical surfactants readily assemble at surfaces, it is not trivial to remove them, making their application non-reversible in practice. Consider how easy it is to dissolve soap in water while washing your hands, yet how difficult it is to remove the soap from the water. Rather than removing surfactant, it is possible to use chemical changes to the surfactant molecules to alter their behavior. There are several examples in literature, but these chemically modified surfactants only result in modest changes in surface tension and in several cases the changes are permanent [30]–[33].

Another way to lower the effective interfacial tension is using electrical potential, which is possible for liquid metals in electrolytic solutions. In the absence of surface oxides, the effective interfacial tension of the metal is enormous, and thus, the metal adopts a spherical shape. The surface oxide can be removed via dissolution by submerging the liquid metal in either a low or high pH solution, such as 1M HCl or 1 M NaOH respectively. In addition, the oxide can be removed electrochemically by applying reducing potentials to the metal (e.g. $-1$ V) in electrolyte [34]. In principle, switching the potential to a sufficiently positive value should deposit oxide electrochemically and the metal should remain spherical due to the large interfacial tension. In fact, in air liquid metal forms a native oxide, a solid skin that encases the metal and prevents the metal from changing shape (as long as there are no forces that exceed the surface yield stress of the oxide [11]). Thus, one might expect electrochemical oxide deposition to create a thin, yet solid skin that would prevent the metal from changing shape. Applying a positive potential (~1 V) in 1M NaOH indeed deposits oxide species, but remarkably it causes an enormous drop in the effective interfacial tension, under certain conditions approaching nearly zero [14]. With a lowered effective interfacial tension, droplets of the metal flatten in response to gravity and under certain conditions, can even form fractals, **Figure 1** [15]. Unlike classical surfactants, oxide species can be removed by applying a reductive potential -1 V, as shown in **Figure 1**.

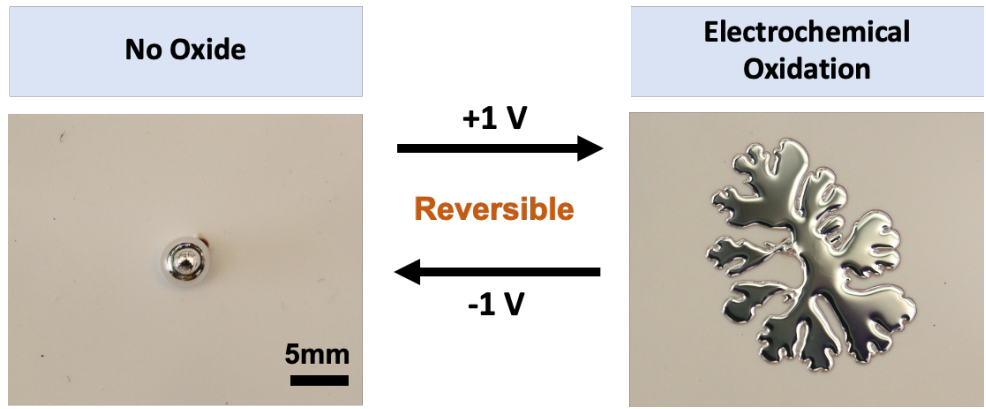

**Figure 1.** EGaIn in 1M NaOH solution. Liquid metals have enormous surface tension and form spherical droplets in the absence of oxide species (left). Oxide species can be removed by applying a reducing potential (here, -1 V) to the metal in electrolyte relative to a counter electrode, copper. However, during electrochemical oxidation (oxidizing potential, +1 V), the electrochemical oxidation lowers the effective interfacial tension to near-zero and gravity drives it to flatten (right). The surprising fractal pattern results from Marangoni forces due to the effective interfacial tension at the outer edge being lower than other regions of the metal.

**Electrochemically modulated effective interfacial tension**

This progress report focuses on the use of electrochemical oxidation to lower the effective interfacial tension of liquid metals. The effect has also been called 'electrochemical capillarity' in the literature [35]. Electrochemically-driven oxidation processes are often called anodization, although we avoid this term because it is usually associated with the more specific case of thickening the surface oxide layer [36]. This method of tuning effective interfacial tension is limited to liquid metals in electrolyte but has the appeal of being tunable and reversible using potential.

The first known observation of this decrease in effective interfacial tension attributed its cause to electrocapillarity [37], and that mistaken association still arises today. Electrocapillarity lowers the effective interfacial tension via an applied electric potential that draws counter ions from the surrounding solution to the surface of a metal [13], [38], [39]. These ions enhance the electrical double layer that forms at the interface between a metal and electrolyte, and act as a capacitor. Whereas interfacial tension forces result from molecular attractions, like-charges within the plane of the double layer create repulsive forces that lower the interfacial tension. The voltage-dependence, to leading order, is described by the electrocapillary equation,

$$\gamma(V) = \gamma_{pzc} - \frac{1}{2}C(V - V_{pzc})^2 \qquad (1)$$

where $\gamma_{pzc}$ is the interfacial tension at the point of zero charge (PZC), $C$ is the capacitance per unit area of the double layer (typically tens of µF/cm²), $V$ is the applied electric potential and $V_{pzc}$ is the potential at which there is zero charge. Equation 1 produces a parabolic shape with the maximal interfacial tension occurring at $V_{pzc}$ because the surface in that state has the least charge (least repulsion). In reality, the electrocapillary curve is only approximately parabolic since the width and symmetry depend on the type/concentration of electrolyte and absorption of ions at the interface [40] Here, it is helpful to think of interfacial tension as the energy cost (per unit area) to create new interface; in that context, the last term in Eq. (1) represents the energy density of a capacitor on a per-unit-area basis. Formally, this equation derives from integration of the Lippmann equation, and is only valid in the absence of reactions [40], [41].

While electrocapillarity is an appealing effect – Eq. (1) suggests that the interfacial tension of a liquid metal could be reversibly tuned to any value, including zero – applied voltages beyond ~1 V cause Faradaic reactions between the metal and electrolyte. Typically, these reactions cause electrolysis of the surrounding water, generating bubbles of oxygen or hydrogen depending on whether the potential applied to the liquid metal is positive (oxidizing) or negative (reducing), respectively. This limits the use of electrocapillarity to low electric potentials. Assuming a typical double layer capacitance of about 27 $\mu$F/cm² [14], an applied potential of approximately 2.5 V above/below $V_{pzc}$ would achieve zero interfacial tension. Yet, this does not happen in reality because such large potentials drive reactions that are not captured by the capacitive effects of

Equation 1 [41]. Consequently, liquid metals remain at relatively high interfacial tensions (several hundreds of mN/m) even under the most effective electrocapillary conditions.

Pure liquid gallium exhibits characteristic electrocapillary behavior, illustrated in **Figure 2** using data from the first known report for gallium [41]. Because gallium is highly reactive and prone to oxidation at modest potentials, the parabolic shape is truncated above 0.8 V. To obtain this plot, the pH of the electrolyte had to be adjusted for each data point to avoid reactions leading to either hydrogen evolution or oxide layer formation. Without this adjustment, a single electrolyte with constant pH would have limited the measurements to a much narrow range of electric potentials. The data points are well-fit by a parabola for which the capacitance was independently measured, likely using electrochemical impedance spectroscopy (EIS). Importantly, the interfacial tension reported in Figure 2 spans from 630 mN/m to as low as 540 mN/m. Although the delta between these values is non-negligible, the metal remains approximately spherical at such enormous absolute values of tension. Thus, electrocapillarity cannot explain the unusual behavior of the metal in Figure 1.

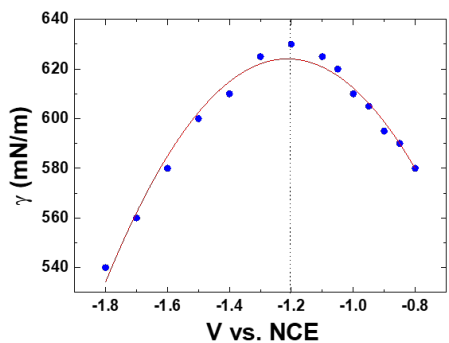

**Figure 2.** Electrocapillary curve of gallium in $x$ KCl + $y$ HCl. Importantly, $x$ and $y$ are not constant: each data point is measured in different pH to avoid Faradaic reactions, which enables electrocapillary behavior over a wider potential window. Note that the tension only decreases from approximately 630 to 540 mN/m using this approach. The blue circles are the interfacial tension measurements from experimental data (reproduced from [41]) and are fit to the electrocapillary equation (Eq. (1)).

In **Figure 2,** a normal calomel electrode (NCE) served as a reference electrode in a three-electrode system in which the liquid metal (here, molten gallium) serves as the working electrode. However, in more recent studies, calomel electrodes (e.g. NCE or Standard Calomel Electrode (SCE)) have widely been replaced by Ag|AgCl|KCl$_{3M}$ reference electrodes for environment reasons. **Figure 3** shows how to convert potentials between measurements made using different reference electrodes. For example, a potential of -1.5 V vs. Ag|AgCl|KCl$_{3M}$ is the same as -1.539 V vs. SCE and -1.575 V vs. NCE and -1.295 V vs. NHE (Normal Hydrogen Electrode), if all other conditions remain the same.

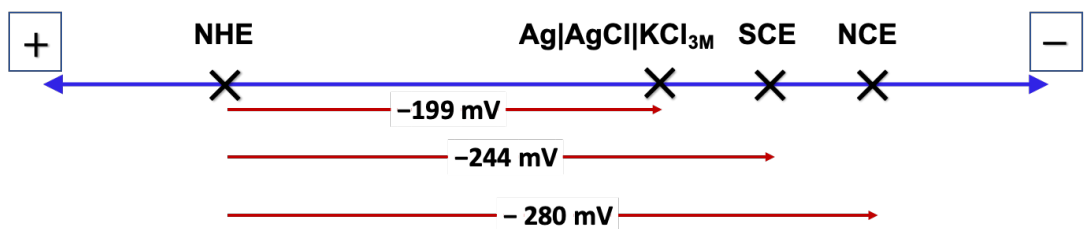

**Figure 3.** Converting potentials from NHE to Ag|AgCl|KCl$_{3M}$, SCE, and NCE reference electrodes.

In the discussion that follows, all measurements were performed in 1M NaOH unless otherwise stated, for which the open circuit potential of EGaIn is approximately -1.5 V (vs. Ag|AgCl|KCl$_{3M}$). All the reported potential values that follow are shifted relative to the open circuit potential (OCP) by adding to the measured value (i.e. $V = 0$ at the OCP) for better understanding for non-electrochemists.

**Comparison between classical electrocapillary behavior and electrochemical oxidation**

Electrocapillarity only modestly lowers interfacial tension due to being relevant only within the small electrochemical range of potentials that avoid Faradaic reactions, as noted previously. However, in some cases (as we discuss here), reactions at the metal surface result in a change of effective interfacial tension.

In 2014, our group quantified the surface activity of EGaIn via surface oxidation [14]. The surface behavior was observed at least once previously (2009) but was attributed to electrocapillarity [41]. **Figure 4a** shows experimental effective interfacial tension data of EGaIn via electrochemical oxidation using sessile droplet measurements. Such measurements determine effective interfacial tension from the sessile shape of a drop. The shape of the drop represents a force balance between effective interfacial tension (unknown, but determined by the shape) and gravity (known as long as density and droplet shape are known) [42]. Using sessile drop-shape analysis it is possible to determine effective interfacial tension versus potential, as plotted in **Figure 4a**. This plot also includes a single magenta-colored (and square-shaped) data point that estimates the near-zero effective interfacial tension of liquid metal at 0.8 V, which will be explained in the following section. We distinguish the magenta-colored point because such a low effective interfacial tension cannot be measured easily by drop shape analysis and we estimated in other ways.

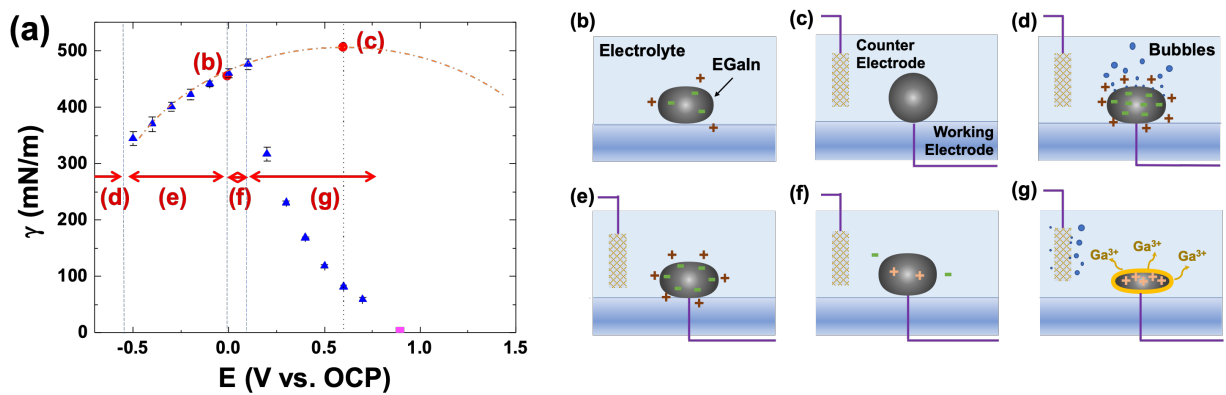

**Figure 4**. (a) Effective interfacial tension (blue triangles) of EGaIn measured in 1M NaOH as a function of potential (E). The magenta square estimates the effective interfacial tension based on the capillary height of the droplet of metal. The trajectory of the blue data intercepts the potential axis near 0.8 V, which is the value where near-zero effective interfacial tension behavior occurs (cf. Figure 1). The red dotted parabola represents the hypothetical interfacial tension if EGaIn followed the electrocapillary Equation 1. The vertical dotted lines are intended to guide the eye. (b-

g) Series of images that illustrate the liquid metal at each point shown in (a). See text for details. A three-electrode system was used for the experiments, but the reference electrode is not depicted in this figure.

This effective interfacial tension is co-plotted with the theoretical electrocapillary as a dotted parabola curve in **Figure 4a** assuming Eq. (1) while using a capacitance of EGaIn based on a constant and typical capacitance associated with an electrical double layer (~28 $\mu F/cm^2$, as measured in reference [14]). EGaIn follows the conventional electrocapillary behavior over a range of potentials (V < 0.05 vs. OCP), however it deviates from electrocapillary theory at V > 0.05 vs. OCP as the surface oxidizes [14]. Electrochemical oxidation of the metal begins to occur near this potential, which lowers the effective interfacial tension significantly. Thus, two key findings: (1) the lowering of effective interfacial tension above ~0.1 V vs. OCP is <u>not</u> due to classic electrocapillarity, and (2) the effective interfacial tension lowers by an unprecedented amount over a small range of potentials for reasons that are still being elucidated.

The series of schematics in **Figure 4b-e** depict the surface state of EGaIn at each potential range. At open circuit potential (V= 0, no external potential applied), EGaIn is presumed to be negatively charged based on its position on the electrocapillary curve (**Figure 4b**). In principle, the maximum interfacial tension should occur at the point of zero charge, in which application of positive potential offsets the negative charge of the metal (**Figure 4c**). From the parabolic fit, we estimate $V_{pzc}$ ~ 0.6 V vs OCP at point (c) in **Figure 4,** but emphasize that this is a hypothetical point predicted by assuming EGaIn follows conventional electrocapillary; in reality, EGaIn electrochemically oxidizes at voltages below $V_{pzc}$.

As shown in **Figure 4a**, interfacial tension of EGaIn follows conventional electrocapillary behavior at cathodic (negative) potential (**Figure 4e**) up to low positive potentials (< 0.05 V vs OCP, **Figure 4f**). At these potentials, the surface should be free of oxide species due to the 1M

NaOH electrolyte. Here, the electrolyte serves two important roles. First, this basic solution dissolves the oxide on the surface to form soluble gallates ($NaGa(OH)_4$). This dissolution of oxide species also competes with oxide deposition during electrochemical oxidation (i.e. at elevated positive potentials). Second, the NaOH ions increase the conductivity of the water and thereby facilitate the rate of electrochemical reactions; this ensures that there is negligible ohmic potential drop in the electrolyte.

In the absence of oxide, applying a more *negative* potential increases the charge density, which reduces interfacial tension by classic electrocapillarity, as shown by the agreement between the data and the fit using Eq. (1) in **Figure 4a**. Below −0.5 V vs. OCP, hydrogen bubbles form on the EGaIn surface and the interfacial tension stops changing with potential (**Figure 4d**). Thus, −0.5 V vs. OCP represents the lower limit for electrocapillarity for these conditions.

In contrast, the behavior at anodic (positive) potentials deviates strongly from classic electrocapillarity above ~0.1 V vs. OCP. Effective interfacial tension drops rapidly as oxide deposits electrochemically on the surface, as shown in **Figure 4g**. At these potentials, EGaIn droplets start to increase in surface area (distorted by gravity) as effective interfacial tension lowers.

**Electrochemical oxidation of liquid metal**

**Figure 5a** illustrates how EGaIn in 1 M NaOH aqueous solution changes shape as a function of electric potential. The shape of the liquid metal is determined by the balance of two forces: effective interfacial tension and gravity. At OCP, the EGaIn droplet resembles a spherical shape due to its high interfacial tension (~500 mN/m) [14]. At 0.5 V vs. OCP, where the effective interfacial tension is about 120 mN/m, the liquid metal flattens and before reaching a new steady-state shape that does not spread further. By applying 1 V vs. OCP, the effective interfacial tension of liquid metal reaches near-zero and becomes unstable: it continues to spread and increase in area until it eventually breaks away from the electrode. In that unstable state, the use of classic drop-shape analysis to measure interfacial tension is no longer possible. Nevertheless, the effective interfacial tension can be estimate based on the so-called capillary height, which assumes that gravitational forces (proportional to the thickness of the metal) balance interfacial forces. Such an estimate provides the magenta data point in **Figure 4a** at 0.8 V.

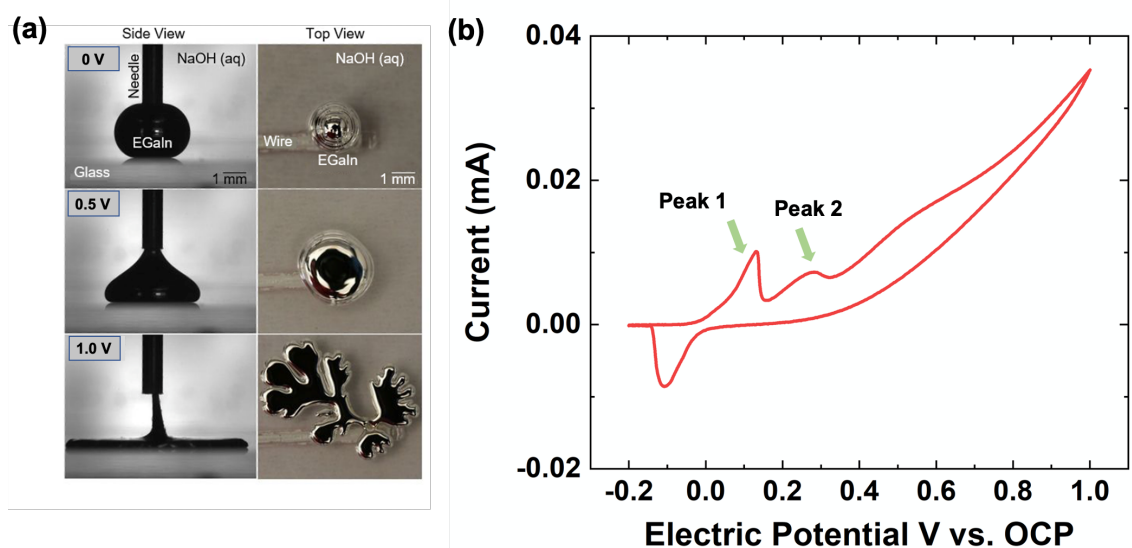

**Figure 5.** The effective interfacial tension (and thus, shape) of a droplet of liquid metal (EGaIn) changes dramatically with respect to electric potential due to electrochemical surface oxidation. (a) A liquid metal droplet is dispensed from the end of a needle. A potential is applied to the liquid metal in 1 M NaOH relative to a counter electrode, which is not visible in this field of view. The change in shape is apparent from the side (left column) and top (right column) views. The effective interfacial tension decreases dramatically as a function of potential and behaves as if the effective interfacial tension is near-zero at ~1 V. Adapted from ref. [14] with permission from PNAS. (b) Cyclic voltammogram of EGaIn (scan rate: 5 V/s) which shows two oxidation peaks (marked green arrows).

Considering the importance of electrochemical oxidation on the effective interfacial tension, we briefly discuss electrochemistry. Cyclic voltammetry (CV) is an electrochemical technique that uses a potentiostat to measure current while sweeping the electric potential with time. In the three-electrode system, the potentiostat measures the potential between the working and reference electrodes. Meanwhile, the potentiostat also measures current between the working and counter electrodes. CV involves scanning the potential linearly in anodic and then cathodic directions to study electrochemical oxidation and reduction reactions, respectively. When potential is scanned in the increasing anodic direction (that is, positive relative to the OCP), it measures an anodic current as a result of oxidation reactions [43]–[45]; an example is shown in **Figure 5b**. In contrast, when potential is scanned in the cathodic direction, it measures a cathodic current as a result of reduction reactions. Anodic (positive) and cathodic (negative) currents depend on the scan rate and the details of the experimental configuration.

As shown in **Figure 5b**, CV measurements on EGaIn exhibits multiple anodic current peaks (oxidation peaks), where each peak, marked by green arrows, represents an electron transfer reaction. Electrochemical oxidation of gallium is complicated due to chemical reactions that can occur in addition to the electrochemical reactions [46]. Given these considerations, we present a voltammogram at fast scan rates (5 V/s) performed on a small EGaIn droplet (0.126 mm$^2$) in **Figure 5b.** A fast scan rate decreases the time window of the experiment, reducing the contribution of chemical reactions relative to electrochemical reactions. Fast scan rates result in two additional effects. First, the location of the peaks in irreversible electrochemical reactions shift as scan rate increases: oxidation peaks shift positively and reduction peaks shift negatively. Second, the electrical current increases with scan rate and consequently the position of the current peak may shift due to increasing potential (IR, or current times voltage) drop. While the precise location of the peaks depends on a number of factors and it is not our intent to go into more detail here, it is notable and important that the two peaks occur within the range of potentials where the effective interfacial tension decreases substantially.

We expect to observe three oxidation peaks, one for each of the three electrons lost by gallium to reach the most stable Ga$^{3+}$ state. By comparing the larger amplitude of electrical current in peak 1 relative to peak 2, it can be reasoned that peak 1 is a two-step electron transfer process (Ga to $Ga^+$ and $Ga^+$ to $Ga^{2+}$). Peak 2 is likely the final step ($Ga^{2+}$ to $Ga^{3+}$) of Ga electrochemical oxidation. It should be noted that $Ga^{2+}$ is highly unstable and therefore can only be detected via fast scan voltammetry, as shown here. At lower scan rates, it chemically converts to Ga and $Ga^{3+}$ via a disproportionation chemical reaction. By comparing the CV (**Figure 5b**) and effective interfacial tension graph (**Figure 4a**), it is noteworthy that the first reaction peak occurs near 0.1 V vs. OCP, which is similar to the potential at which effective interfacial tension anomalously drops (**Figure 4a).** Taken in sum, we include this brief discussion of electrochemistry to emphasize that the drop in effective interfacial tension coincides with electrochemical oxidation.

**Electrochemical oxidation mechanism**

Gallium-based alloys rapidly form an oxide layer (~1-3nm thick) in air [11], [47]–[51] that passivates the surface in the presence of oxygen [52]. The surface oxide of EGaIn in the air is composed primarily of gallium oxide (Ga2O3), despite the presence of indium in the bulk [11]. The oxide is passivating, which means it does not get thicker with time, although in wet environments with neutral pH, the oxide changes to GaOOH and becomes less passivating [53]. Although oxides formed in air likely lower the effective interfacial tension of liquid metal relative to bare metal, the liquid does not adopt a surface-minimizing shape because those oxides are sufficiently rigid. For example, prior studies have reported modest decreases in the surface tension of molten metals (from 515 mN/m to 495 mN/m [54]) as a function of partial pressure of oxygen [54], [55]. It is difficult to know the full extent to which native oxides lower tension. Considering that electrochemical oxidation can monotonically lower tension as a function of potential, it seems likely the electrochemical mechanism for lowering the effective interfacial tension goes beyond what is possible with native oxides alone.

Regardless of how it is formed, an oxide layer can be removed easily by using acid, base, strong reductants, or electrochemistry [56]. The oxide is said to be *amphoteric* since it can be removed with either acid or base. The use of HCl solution converts $Ga_2O_3$ to gallium trichloride ($GaCl_3$) and NaOH solution forms sodium tetrahydroxogallate (III) ($Na[Ga(OH)_4]$) [57]. Upon removing the oxide, liquid gallium alloys bead up into spherical shapes due to their large interfacial tension. The stable gallium oxidation state is $Ga^{3+}$, and thus possible gallium oxide species in solution are $Ga(OH)^{2+}$, $GaO^+$, $GaO_2^-$, $H_2GaO_3^-$, $HGaO_3^{2-}$, $GaO_3^{3-}$ and $Ga(OH)_4^-$ depending on pH of the solution [46]. These species are predicted to be the most favorable thermodynamically, but this does not guarantee they are the same species that form dynamically on the surface during electrochemical reactions.

In a basic solution, studies suggest gallium forms oxides or hydroxide films by electrochemical oxidation on the surface which consist of GaO(OH) (gallium oxyhydroxide), $Ga_2O_3$ (gallium oxide), or $Ga(OH)_3$ (gallium hydroxide) [58]–[61]. However, the chemical composition of the oxide created during electrochemical oxidation remains uncertain [46]. Identifying the chemical composition in basic solution is challenging for several reasons. First, electrochemical oxidation of gallium is a complex mechanism including coupled chemical and electrochemical reactions that may change as a function of applied electric potential. Second, the biggest experimental challenge is that the oxidation products on the surface of the metal rapidly dissolve in basic electrolyte [60]. The dissolution makes it hard to identify the surface composition during active electrochemical oxidation, the process of most interest.

Given that decreases in effective interfacial tension only occur at potentials where oxidation is favorable (and confirmed by the observation of electrochemical current), it is clear that electrochemical oxidation is responsible for lowering effective interfacial tension. In our initial studies, we speculated that the oxide acts as a surfactant that separates the metal from the electrolyte, but this does not fully explain how the effective interfacial tension continues to drop with potential beyond 0.1 V vs OCP [14]. In subsequent work [11], we argued that at sufficiently large potentials, large enough to drive electrochemical oxidation through surface oxide species, a second mechanism helps further lower the effective interfacial tension. The following section elaborates on additional observations and measurements that help clarify the behavior of liquid metal during this process.

**Surface activity of EGaIn during electrochemical oxidation**

Studies on the surface activity of sessile droplets or flowing streams of EGaIn as a function of potential provide insights into the physical impact of electrochemical oxidation on the behavior of the liquid metal [15], [62]. In the case of the sessile drop (resting on a surface), measurements of effective interfacial tension are possible only when the initial diameter of the drop is large enough that gravitational force is on the same order of magnitude as the effective interfacial tension. This dimension is called the capillary length: $L = \left(\frac{\gamma}{\rho g}\right)^{1/2}$, where $L$ is the characteristic capillary length, $\gamma$ is the effective interfacial tension of the EGaIn, $\rho$ is the density of the EGaIn, and $g$ is the acceleration due to gravity. The capillary length of bare metal is ~3 mm. Using sessile droplets whose diameter is near the capillary length ensures they get distorted by gravity, which also aids observing physical changes to the liquid metal (**Figure 5**).

Changes in tension can also affect the morphology of liquid metal pumped from a nozzle into an electrolyte. The shape of a drop hanging from a nozzle arises from the combined effects of gravitational deformation and restorative forces from the effective interfacial tension. The

Worthington number, $W_0 = \frac{\Delta \rho g V_d}{\pi \gamma D_n}$, (where $V_d$ is the volume of the EGaIn droplet and $D_n$ is the diameter of the needle) is a dimensionless number shows the ratio between gravitational deformation and the effective interfacial tension. When $W_0 \sim 1$ is considered as an accurate and valid measurement [63].

**Figure 6** shows the phase diagram of the morphological regimes of both a sessile droplet and a flowing stream of EGaIn as a function of electric potential. At ~ 0.15 V vs OCP, the electrochemical oxidation of the metal causes the effective interfacial tension to decrease [14], causing the EGaIn sessile droplets to *distort* (in a previous publication, we called this Regime A [15]), as shown in **Figure 6a**. In the case of a flowing stream of EGaIn, lowering the effective interfacial tension causes *drops* to fall from the nozzle more frequently at smaller volumes until ~ 0.8V vs OCP, as shown in **Figure 6b**. At the same potential where sessile droplet starts to *spread* into fractals (Regime B), the EGaIn stream transitions to smooth *wires*, suggesting the effective interfacial tension is near-zero in both droplets and streams. As a larger positive potential is applied (> 1.8V vs OCP), the oxide thickens. The thickened oxide is visually apparent due to the formerly shiny surface appearing matte and the current dropping precipitously. Consequently, the EGaIn droplet *stops spreading* and *retracts* (Regimes C & D) [15]. In a similar potential range, EGaIn streams (**Figure 6b**) form irregular, drop-like *blobs* as potential increases. Additional morphologies can occur at particular conditions but are omitted here for simplicity.

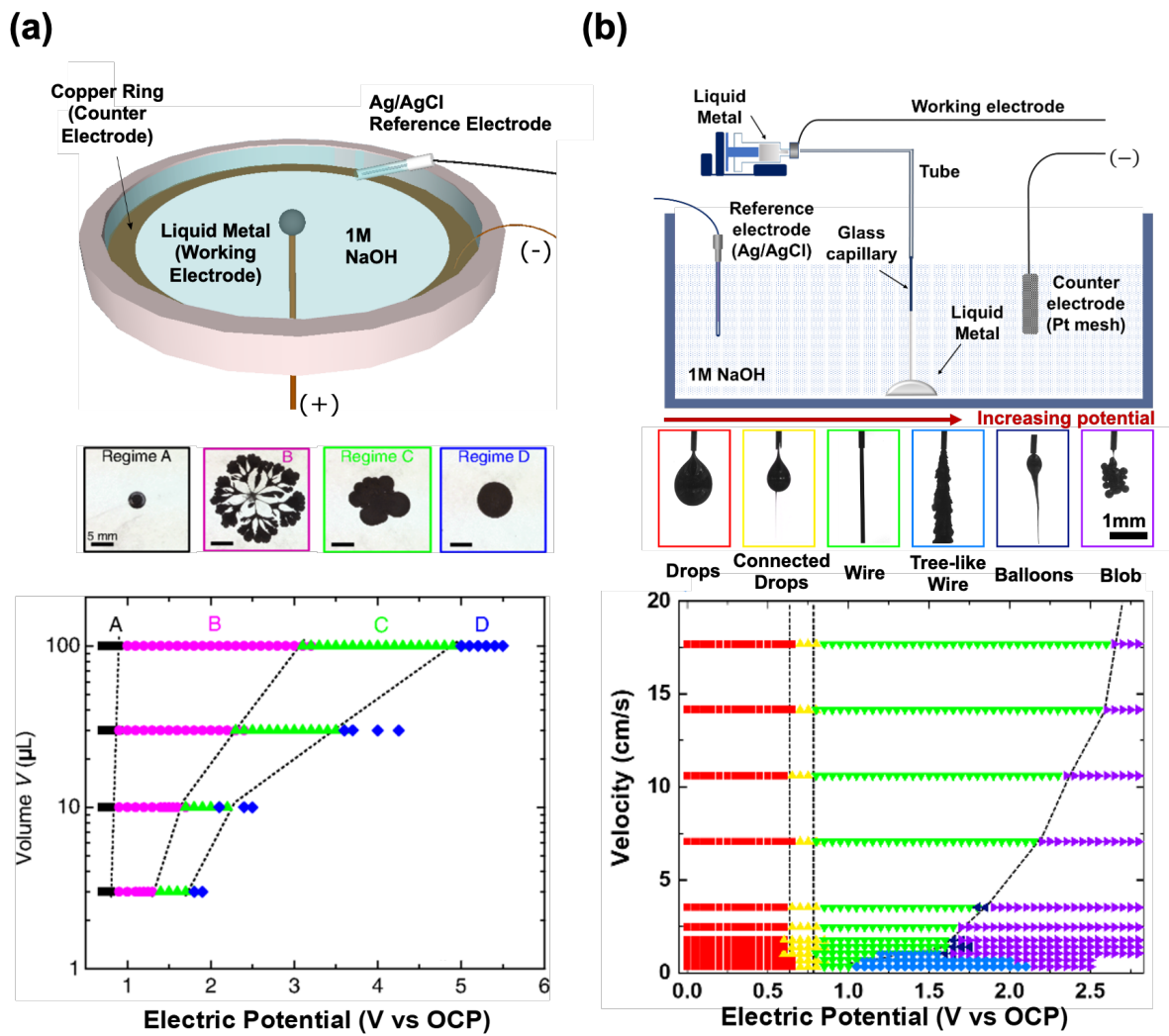

**Figure 6.** Morphological behavior of EGaIn as a droplet and stream plotted versus electric potential. 1M NaOH was used as electrolyte in both experiments. (a) Potential can be applied to a sessile drop in a flat dish (top schematic). The row of images shows different morphologies observed by looking top down in the dish. Four different regimes emerge as a function of volume of the sessile drop and electric potential. (b) Potential can be applied to a stream of liquid metal emerging from a nozzle (top schematic). The row of images shows six different morphologies result as a function of flow velocity and electric potential. The color schemes between (a) and (b) have no correlation and are simply adapted in their original form with permission from (a) PRL [15] and (b) PNAS [64].

Despite the differences in the two experimental setups, the changes in morphology correlate with potential, whether for a sessile drop (**Figure 6a**) or for a flowing stream (**Figure 6b**). This observation confirms the importance of potential: (1) *Regime A* for sessile drops corresponds to the *drop* regime for fluid streams in which effective interfacial tension is finite and decreases with potential, (2) *Regime B* corresponds to the *wire* regime in which effective interfacial tension is near zero and sessile droplets remarkably form fractals and fluid streams remarkably form cylindrical wires, and (3-4) *Regimes C and D* correspond to the *Balloon* and *Blob* regime in which case the surface forms a thicker oxide that obstructs shape change at high potentials.

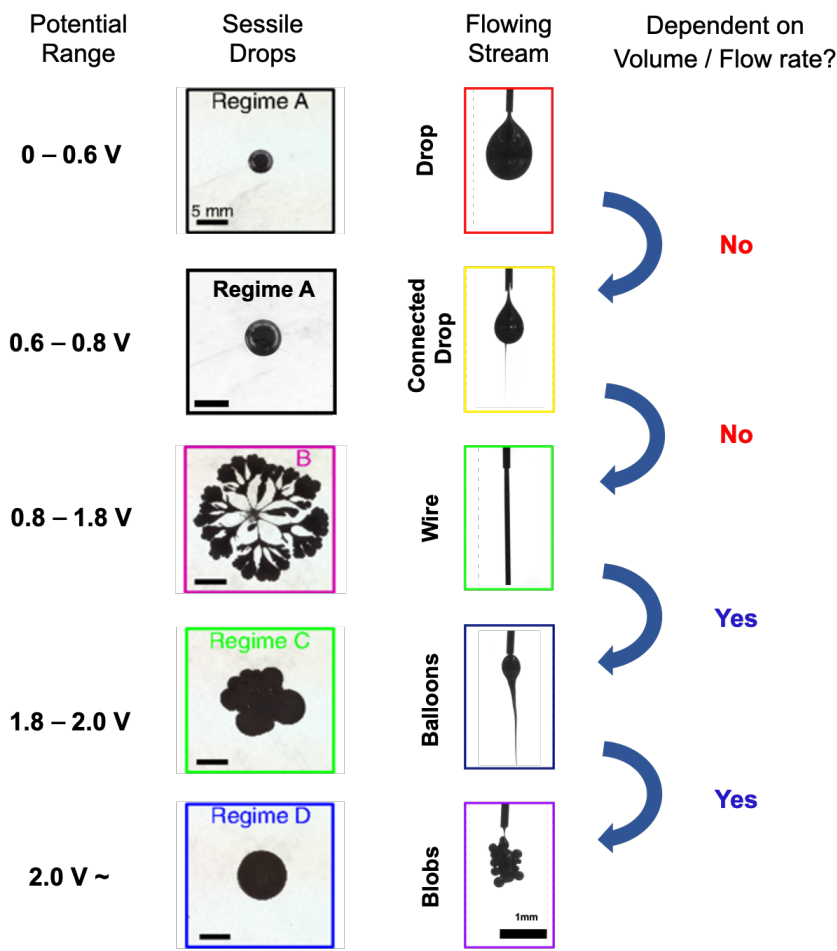

**Figure 7.** A side-by-side comparison of morphological behavior between sessile droplets and flowing streams of EGaIn at different electric potential in 1 M NaOH. The potentials at which the transitions occur are independent of the volume of the sessile drops (or likewise, flow rate of streams) except at the transitions that occur at the highest potentials. Thus, these latter transitions are likely mass-transfer limited.

For both sessile drops and flowing streams, the transition from Regime A to B (in the case of sessile drops) and drop to wire regime (for flowing streams) occurs at the same relatively low potential (~0.8 V vs OCP). The transitions are also independent of the volume of the sessile drop and velocity of the liquid stream (**Figure 7)**, which suggest that the morphology of the metal and effective interfacial tension is driven primarily by electrochemical reactions. However, the transition to the final Regimes depends on drop size or flow rate, which suggests an additional transport-dependence that has not been fully elucidated.

**Morphological behaviors at higher potentials**

Recall that at higher electric potentials, the surface oxide thickens and provides a mechanical shell on the surface. Unlike other morphological transitions, this transition occurs at potentials that depend drop size (in the case of sessile drops) and flow rate (in the case of fluid streams). We have previously reasoned that these dependencies arise from the choice of NaOH as the solution [14], [15], [62], [65], because NaOH serves as an electrolyte that can dissolve oxide species as they are created. The dissolution of gallium oxide proceeds through the reaction,

$$\text{Ga}_2\text{O}_3 + 2\text{ NaOH} + 3\text{ H}_2\text{O} \rightarrow 2\text{ Na}[\text{Ga(OH)}_4] \qquad (2)$$

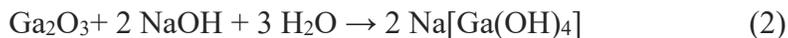

Thus, applying a potential drives oxide formation, which competes with oxide dissolution. We reason that the thick oxide layer forms when the rate of deposition exceeds dissolution based on the sudden drop in current. This has implications for morphological transitions that involve this thickening oxide. We also note a complicating factor: the dissolved form of Ga is a gallate species—Ga(OH)$_4$—which has a negative charge and therefore will be electrostatically attracted to the oxide/electrolyte interface, thereby obstructing or slowing the ability of OH- to freely reach the oxide. A similar effect has been observed in Al during anodization [66].

At higher potentials, the morphological behavior of the metal depends on both the potential and the area of the sessile drop (or alternatively, the velocity of the flowing stream) (**Figure 7**). For example, the transition from Regime B to C/D (or wire to blob, in the case of a flowing stream) requires a larger potential with increasing volume, suggesting a transport limit. Namely, the deposition rate of oxide has to exceed the rate of dissolution. It seems likely that forming a thick layer requires a higher potential as the area (or rate of area generated, in the case of a flowing stream) increases, but this behavior warrants further study. The supporting information of reference [64] contains additional discussion and analysis of this issue.

**Morphological behaviors at near-zero effective interfacial tension**

The most remarkable aspect of liquid meal electrochemical oxidation is the near-zero effective interfacial tension behavior. At ~ 0.8 V vs OCP, sessile drops form fractal shapes (Regime B) and liquid streams form wires as shown in **Figure 7**, suggesting the effective interfacial tension is near zero. Therefore, effective interfacial tension in these regimes cannot be well-measured using either sessile or pendant shape analysis [42]. Instead, we estimate the effective interfacial tension of liquid metal by using the areal footprint of a fractal shape to determine the capillary height. The fractal shapes are thought to arise from Marangoni instabilities due to gradients in effective interfacial tension [67]. These fractals increase in area over time, suggesting the effective interfacial tension is remarkably low ($\lesssim$ 0.4 mN/m, three orders of magnitude lower than the interfacial tension of the bare liquid metal) [15]. It is possible the effective interfacial tension is even lower than these values considering the metal never reaches steady-state and continues to increase its area when it fragments.

In the case of a liquid metal stream, the near-zero surface tension allows the formation free-falling cylindrical streams which we call *wires* because of their high aspect ratio and metallic conductivity. Remarkably, these liquid metal wires do not breakup via the Rayleigh-Plateau instability, driven by the destabilizing effects of interfacial tension [68]–[70], instead remaining stable to a height:diameter aspect ratio of at least > 6400, two orders of magnitude higher than similar streams of water [71]. (This is a lower bound on the aspect ratio, as the 64 cm long wire spanned the entire length of the container in which it was created.) The upper limit on the effective interfacial tension of liquid metal wire is $1.5 \times 10^{-8}$ mN/m, based on a Rayleigh-Plateau instability scaling relationship, although such an estimate neglects any visco-elastic effects of the surface species that likely help stabilize such streams.

These reports suggest the effective interfacial tension may indeed be zero. One might expect a liquid with zero effective interfacial tension to simply dissolve or provide no resistance to spreading, yet the presence of a thin oxide that separates the metal from the electrolyte limits it its mobility. The importance of an outer shell is emphasized by our observation of string-like behavior over short time scales, in which such wires can be draped across surfaces and move upward when tugged suddenly: this behavior is not characteristic of a purely viscous fluid.

We have established throughout this report that electrochemical oxidation lowers tension, but we have not clarified *how* it lowers tension since this remains an ongoing area of inquiry. There seem to be at least two possible explanations. First, oxides species should exhibit surfactant-like behavior for lowering interfacial tension. The composition (including charge) and concentration of oxide species may vary with potential, thereby explaining the dependency of effective tension with potential. Second, we speculated previously that tension could lower, in part, due to compressive stress arising from electrochemical oxidation. **Figure 8** provides evidence for this mechanism by contrasting the behavior of liquid metal in *Regime B* in electrolytes of different pH. Unlike NaOH solutions, the NaF does not dissolve the oxide and the liquid metal thereby forms a thick oxide that suppresses fingering instabilities (**Figure 8b**). The thick oxide is observable by the naked eye, since the surface transitions from shiny (thin oxide layer, bare metal) to matte (rough). Yet, the liquid metal forms localized protrusions that emerge from this crusty shell, as shown in **Figure 8b.** We reason that these protrusions provide evidence of oxidative (compressive) stress arising from electrochemical oxidation [15]. In addition, we note that at higher potentials (~2 V), the current drops rapidly and fractals recede back to a flattened droplet, as shown in Figure 7. This observation suggests the fractals only occur under the dynamic conditions of electrochemical oxidation; in other words, having oxide species on the surface is insufficient to create fractals.

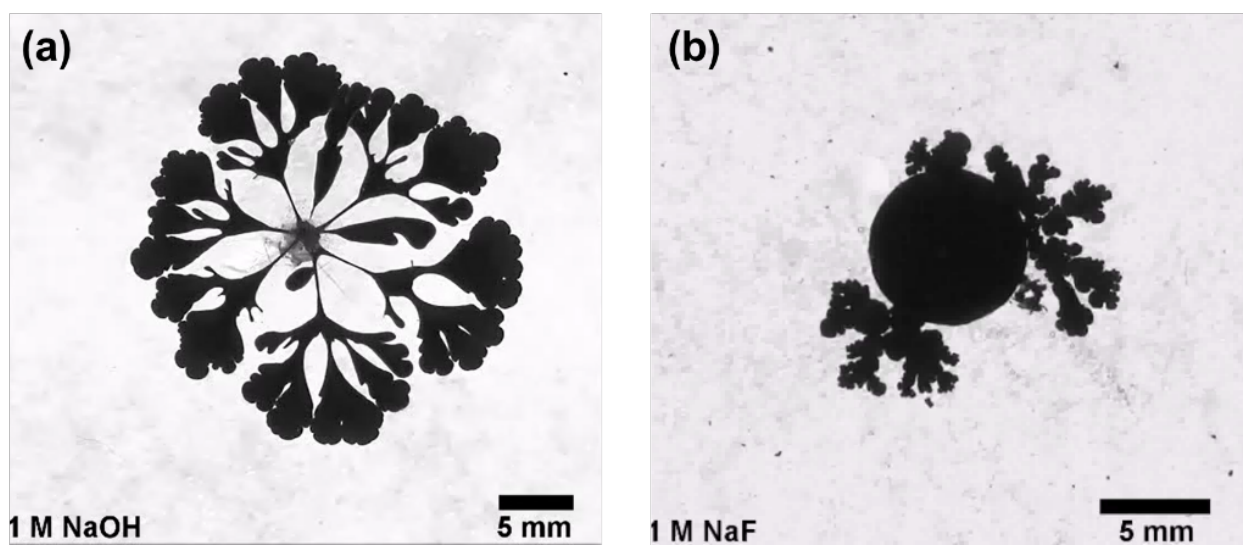

**Figure 8.** Top down photographs of liquid metal when the effective interfacial tension is "near zero". Note that these still images are taken from dynamic movies in which the surface is spreading outward.[15] These experiments use the apparatus shown in Figure 6a. (a) An EGaIn sessile droplet forms dynamic fractals in 1M NaOH. Although not apparent here, the surface of the metal remains shiny throughout the experiment due to the simultaneous electrochemical oxidation and oxide dissolution in NaOH. (b) EGaIn droplet does not spread uniformly in 1M NaF due to the formation of a thicker shell apparent by eye. Yet, it still forms protrusions that allow the metal to spread outward, providing evidence that electrochemical oxidation can create compressive stresses that push the surface outward. 1.8 V and 1.5 V vs. OCP were applied, respectively.

A number of studies suggest stresses can arise during the anodization of aluminum, which despite being solid, is of interest because it lies directly above Ga in the periodic table and therefore has many similar properties including the formation of an $Al_2O_3$ native oxide. Studies of anodization of Al in NaOH provide several interesting insights, including reports of compressive stress generation during anodization. Compressive stress would oppose effective interfacial tension and help explain how the effective interfacial tension of Ga lowers with potential.

Compressive stress in Al anodization may occur through several mechanisms. One possible mechanism is electrostriction, a property of all electrical non-conductors, or dielectrics, that causes them to change their shape under the application of an electric field. The magnitude of the electrostriction effect increases with potential during oxide growth [72], [73]. Large electric fields applied during anodic oxidation cause stresses resulting in passive film breakdown via pitting corrosion [73], [74]. The formation of an anodic oxide film on Al is also influenced by internal stress in the metal, thermal stress, or stress induced by dissolution [75]. Given the fact that Al shares many similar properties with Ga, electrochemical oxidation in EGaIn also likely generates compressive stress that varies as a function of electric potential. For example, a study of solid metallic gallium anodized under different conditions suggests stress at the metal−oxide interface [76], which can arise due to the volume expansion from formation of anodic alumina oxide [77].

**Applications**

In this progress report, we focus on electrochemical oxidation to manipulate the effective interfacial tension of liquid metals from ~ 500 mN/m to ~ 0 mN/m [15], [64]. This technique can be used to control the movement of liquid metal, allowing change in the shape of antennas [78]–[87] and completion or breaking of circuits [21], [88]–[92]. Beyond these, there are several other methods to manipulate the shape and position of liquid metals. In addition to the surface oxidation described in this progress report, liquid metal can be controlled by mechanochemistry [65], [93], pressure [94], ionic gradient [95], ultra-sound field [96], electric field [97]–[100], magnetic field [101]–[104], and electrochemistry [105], [15], [88], [87], [106]–[110], [65]. **Figure 9** shows a variety of liquid metal applications, using a variety of methods to control the shape and the movements: this is a rich framework in which to develop new devices and techniques.

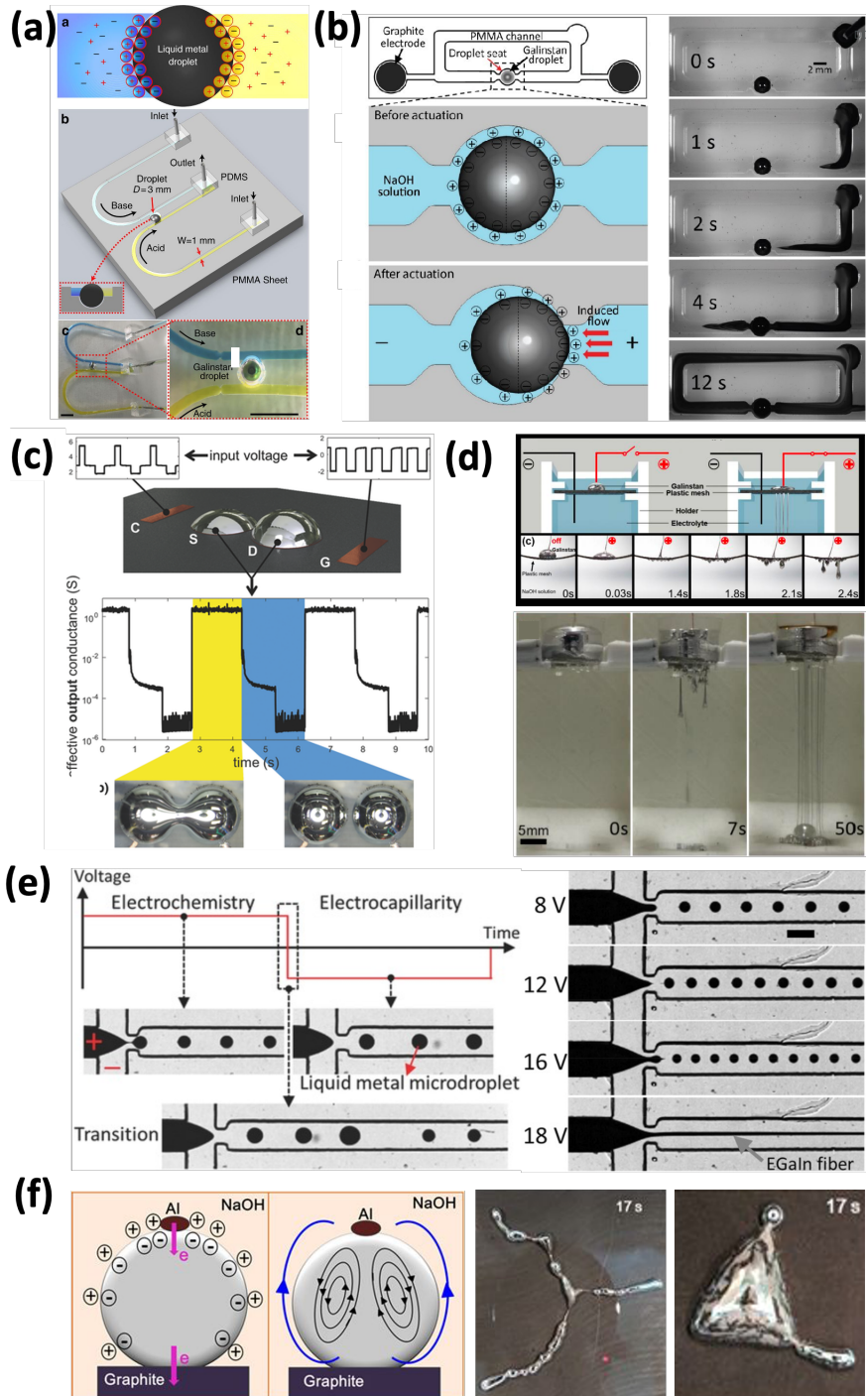

**Figure 9.** Applications that utilize interfacial effects of liquid metal. (a) Liquid metal actuator using pH and ionic concentration gradients [95]. (b) Liquid metal pump controlled by electrocapillarity [17]. (c) Liquid metal transistor using electrocapillarity [111]. (d) Liquid metal penetrating through macro- and microporous materials by applying a voltage [112]. (e) Electrical control of droplet size and rate of liquid metal in microchannel [113]. (f) Surface effect of liquid metal on graphite by adding Al [114].

Starting from the upper left: gradients in pH and ion-concentration can cause interfacial tension gradients that propel liquid metal droplets as a result of electrocapillary effects [95]. As illustrated in **Figure 9a,** the liquid metal droplet moves from the HCl side (high surface tension) towards the NaOH (low surface tension) electrolyte. Supplying different pH electrolytes (HCl and NaOH) on each side of a drop of metal causes electrical double layer (capacitance) imbalance (Eq. (1)), resulting in interfacial tension gradients that can cause the metal to move.

Liquid metal can also be utilized for small-scale pumps using continuous electrowetting (CEW), which is a result of the electrocapillary effect. As shown in **Figure 9b**, when voltage is applied between two graphite electrodes, a potential gradient is generated across a droplet of metal. Gradients in potential cause gradients in interfacial tension, described by the electrocapillary equation (Eq. (1)). Normally, such gradients in interfacial tension cause liquid metal droplets to move. Here, the authors confined the liquid metal within a chamber. Since liquid metal is confined, the potential gradient instead causes the surrounding electrolyte to flow. By changing the frequency and the magnitude of applied voltage, the pumping flow rate can be optimized. Thus, the liquid metal droplet acts as a microfluidic pump.

**Figure 9c** demonstrates a field-controlled electrical switch using liquid metal: two droplets of liquid metal can bridge together (conduct) and break apart (insulate) to form a liquid transistor [111]. The droplets wet Cu pads and therefore it only can deform in response to electric fields applied using two external electrodes. By manipulating gradients in the effective interfacial tension, the droplets coalesce with a neighboring droplet. Upon returning the metal to a state of high interfacial tension, the droplets break apart.

Surface oxidation reduces effective interfacial tension of liquid metal to near-zero, allowing the liquid metal to penetrate through macro/microporous materials (**Figure 9d**). This example demonstrates that extremely low effective interfacial tension and gravity cause the liquid metal to penetrate paper or other porous materials, and fall through as wires on the other side [112].

**Figure 9e** demonstrates liquid metal in a flow-focusing microfluidic system, generating liquid metal droplets of various sizes and controlling production frequencies via electrocapillarity and electrochemistry [113]. Such a microfluidic device consists of two immiscible streams forced through a constriction that breaks one of the streams into droplets (discrete phase) surrounded by a continuous phase. Changing the flow rate of the continuous or discrete phase liquids is a way to control microdroplet diameter. The diameter of the droplets can be monodisperse once the system stabilizes at steady state. However, this process requires a relatively long time to reach steady state, during which time there is little control over the droplet size. In contrast, applying potential to the liquid metal allows the drop size to be tuned in real time by modulating the interfacial tension (lower tension forms smaller droplets).

**Figure 9f** shows an aluminum pellet contacting liquid metal on graphite, which induces the surface charge gradient across the droplet resembling amoeba like behavior [114]. This is attributed to galvanic interaction between the Al and the graphite substrate in the NaOH solution. Therefore, the shape of liquid metal can be controlled by the amount of the aluminum pellet contacting the liquid metal.

Beyond the examples shown in **Figure 9**, a recent study reported electrochemically enabled embedded 3D printing (3e-3DP) for creating freestanding 2D and 3D gallium wire-like structures at large scales [115]. By injecting the liquid metal into hydrogels while applying electric potential allows liquid metal to maintain the freestanding shape. After an enhanced solidification process and the removal of hydrogel, various freestanding 2D and 3D wire-like structures can be obtained, which can be potentially used in flexible, stretchable, and self-healable electronics and devices.

**Challenges and opportunities**

This progress report describes the ability to reversibly tune effective interfacial tension of metals from ~ 500 mN/m to ~ 0 mN/m using electrochemical oxidation [15], [64]. This recent discovery suggests that electrochemical oxidation of liquid gallium and gallium alloys can decrease the effective interfacial tension and can be used for various applications for manipulating liquid metals.

Despite the interest in electrochemical oxidation of gallium alloys, there are many remaining challenges. The most obvious challenge is to better understand the mechanism. Here, we highlighted that the oxide likely acts like a surfactant to lower the surface tension, but a second mechanism, likely compressive stress from electrochemical oxidation, may explain the continued drop in effective interfacial tension as a function of potential. Likewise, the change in behavior when the oxide thickens and the current drops precipitously also needs to be studied. Alternatively, anomalous electrocapillary behavior (that is, charged oxide species that exceed the capacitance of electrical double layers) may also contribute.

Understanding the chemical species at the interface is difficult since the process is dynamic in two ways: (1) the liquid metal, and its interface, moves when the effective interfacial tension is near-zero, and (2) the oxide species dissolve off the surface rapidly when the potential returns to zero. Consequently, dissolution rate and thickness of oxide are difficult to measure because electrochemical oxidation and dissolution occur simultaneously. In terms of electrolyte, questions remain about the role of ions and pH (as evident by the morphology of the metal in the different electrolytes in **Figure 8**) [56]. Likewise, the electrochemistry of oxidation needs further elucidation using electrochemical techniques Finally, there remain opportunities to exploit this interesting phenomenon for new applications based on the ability to achieve huge changes in effective interfacial tension using low potentials.

**Conclusion and outlook**

Ga-based alloys have many potential applications due to their useful combination of metallic and fluidic properties at room temperature. In this progress report, we focus on the role of electrochemical surface oxidation of liquid metal. Surface oxidation can significantly decrease effective interfacial tension to a greater extent than using either surfactants or electrocapillarity. Although this method only works within an electrolyte solution, it opens possibilities for Ga-based alloys in various applications by increasing or decreasing effective interfacial tension over an enormous range using only +/-1V. Despite substantial efforts, many challenges and questions remain regarding this fascinating mechanism for modulating effective interfacial tension.